\documentstyle[12pt,aasms,epsfig]{article}

\input epsf

\newcommand{\etal}{{\it et al.\thinspace}}

\begin{document}
\let\typeset\relax

\title{UNIFIED ELECTRONIC RECOMBINATION OF Ne-LIKE Fe~XVII: 
IMPLICATIONS FOR MODELING X-RAY PLASMAS}

\author{Anil K. Pradhan and Sultana N. Nahar}
\affil{Department of Astronomy, The Ohio State University, Columbus, OH
43210}
\author{Hong Lin Zhang}
\affil{MS F663, Los Alamos National Laboratory, Los Alamos, NM 87545}

\begin{abstract}

 Unified recombination cross sections and rates are computed for
(e~+~Fe~XVIII) $\longrightarrow$ Fe~XVII including non-resonant and
resonant (radiative and di-electronic recombination, RR and DR)
processes in an ab initio manner with relativistic fine structure.
The highly resolved theoretical cross sections exhibit 
considerably more resonance structures than
observed in the heavy-ion storage ring measurements at Heidelberg,
Germany. Nonetheless, the detailed resonance complexes
agree well with experiment, and the unified rates agree with
the sum of experimentally derived DR and theoretical RR rates
to $\sim$ 20\%, within experimental or theoretical uncertainties.
Theoretical results may provide estimates of
field ionization of rydberg levels close to the DR peak, and
non-resonant background contributions, particularly close to the RR peak as E
$\rightarrow$ 0. More generally, the unified results avoid the physical and
practical problems in astrophysical models inherent in the separation of 
electronic recombination into RR and DR on the one hand, and further 
subdivision into low-energy $\Delta n = 0$ DR and high-energy 
$\Delta n > 0$ DR in photoionized and collisionally ionized X-ray plasmas.

\end{abstract}




\section{INTRODUCTION}

The {\it Chandra
X-ray Observatory} and the {\it X-ray Multi-Mirror Mission - Newton}
 are observing a variety of sources (Canizares \etal 2000, Predehl \etal
2000) with wide-ranging plasma conditions. The analysis requires 
astrophysical models 
(e.g. Kallman 1995, Brickhouse \etal 1995) whose accuracy depend on
the atomic cross sections for collisional and radiative processes such
as electron impact excitation, photoionization and recombination.
K- and L-shell iron ions are among the most prominent atomic species in
many sources such as active galactic nuclei (Ogle \etal 2000), 
stellar coronae and winds (Schulz \etal 2000), and
cooling flows in clusters of galaxies (Fabian \etal 2000). Excitation,
photoionization, and recombination of Ne-like 
Fe~XVII are of particular interest as this ion is a strong X-ray radiator owing
to a multitude of L-shell excitations at T $<$ 1 keV.

 Although most of these atomic parameters are obtained theoretically,
those need to be benchmarked against available experimental measurements.
In recent years there has been considerable progress in both theoretical and
experimental methods. High resolution electron-ion
recombination measurements on ion storage rings show very detailed
resonance structures in the low energy region usually accessible in 
experiments (e.g. Mannervik \etal 1997, Kilgus \etal 1992). Recently,
such measurements have been done for
(e~+~Fe~XVIII) $\rightarrow$ Fe~XVII (Savin \etal 1997,1999) 
in the low energy region dominated by $\Delta$ n = 0 resonances.
The experimental 
results naturally measured the combined non-resonant and resonant
contributions. These were then processed
to separate the background and extract the resonance contributions (DR)
by fitting with an experimentally deduced beam shape function.
Savin \etal find that their inferred DR rates from low-energy
measurements differ from previous theoretical calculations by up to a
factor of 2 or more. Their best agreeement of $\approx$ 30\% 
is with multi-configuration Dirac-Fock (MCDF) and Breit-Pauli (MCBP) 
calculations in the isolated resonance approximation, assuming
autoionization and the radiative probabilities independent of each
other; detailed cross sections are not calculated. 
The comparison of derived DR rates from individual
resonances (or blends) showed varying levels of agreement.

 As the (e~+~ion) recombination  is unified in nature, it is theoretically
desirable to consider the non-resonant and resonant processes (RR and
DR) together. A unified theoretical formulation has been developed 
(e.g. Nahar and pradhan 1994, NP94; Zhang \etal 1999, Z99), including
 relativistic
fine structure (Zhang and Pradhan 1997), 
and used to compute cross sections and rates for many
atomic systems, such as the K-shell systems C~IV - C~V and Fe~XXIV - Fe~XXV
of interest in X-ray spectroscopy (Nahar \etal 2000a,b; N99a,b). 
The unified
results may be directly compared with experimental results, without the
need to separate RR and DR. In this {\it Letter}  we present
new results for Fe~XVII and show that not only are the unified cross
sections and rates in very good
agreement with experiment, but may also be used to study important
physical effects such as ionization of
high-rydberg bound and autoionizing levels. More generally, the results
demonstrate that the unified calculations avoid the basic inconsistency
and incompleteness of photoionization and recombination data for
modeling of laboratory and astrophysical plasma sources.

\section{THEORY AND COMPUTATIONS}

 From a quantum mechanical point of view photoionization and
recombination may be treated in a self-consistent manner by considering
the same coupled eigenfunction expansion for the core (photoionized or
recombining) ion. For Fe~XVII we write, 

\begin{equation}
\Psi(E; e + Fe~XVIII) = \sum_{i} \chi_{i}(Fe~XVIII)\theta_{i}(e) + 
\sum_{j} c_{j} \Phi_{j}(Fe~XVII),
\end{equation}
where the $\Psi$ denote both the bound (E $<$ 0) and the continuum (E $>$ 0) 
states of Fe~XVII, expanded in terms of the core ion eigenfunctions 
$\chi_i$(Fe~XVIII); the $\Phi_j$ are correlation functions.
The close coupling approximation, based on the
efficient R-matrix method 
(Burke \etal 1971), and its relativistic
Breit-Pauli extension (Scott and Taylor 1981), enables a solution for
the total $\Psi$, with a suitable expansion over the $\chi_i$.
The Breit-Pauli R-matrix (BPRM) method has been
extensively employed for electron impact excitation under the
Iron Project (Hummer \etal 1993, Berrington \etal 1995). 
The extention of the BPRM formulation to unified electronic
recombination (e.g. Z99, N99a,b), and theoretically self-consistent 
calculations of photoionization and recombination is sketched below.

 Resonant and non-resonant electronic recombination takes place into
an infinite number of bound levels of the (e~+~ion) system. These are
divided into two groups: (A) the low-n (n $\leq$ n$_o \approx$ 10) levels, 
considered via
detailed close coupling calculations for photorecombination, with
highly resolved delineation of autoionizing resonances, and (B) the
high-n (n$_o \ge n \leq \infty$) recombining levels via DR,
neglecting the background. In previous works (e.g. Z99)
 it has been shown that
the energy region corresponding to (B), below threshold for DR,
the non-resonant contribution is negligible.
The DR cross sections converge on to the
electron impact excitation cross section
at threshold (n$\rightarrow \infty$, as
required by unitarity, i.e. conservation of photon and electron fluxes.
This theoretical limit is an important check on the calculations, and
enables a determination of field ionization of rydberg levels of
resonances contributing to DR.
 
 Complete details of the extensive BPRM calculations for
photoionization and recombination of Fe~XVII
 will be presented elsewhere.  The multi-configuration target
eigenfunctions $\chi_i$(Fe~XVIII) (Eq. 1) 
are obtained from an atomic structure
calculation with 5 spectroscopic configurations 
$2s2p^5, 2s2p^6, 2s^22p^4 \ 3s, 3p, 3d$,
and a number of correlation configurations, optimized over the resulting
60 fine structure levels using the program SUPERSTRUCTURE (Eissner \etal 1974). 
In the low energy region of interest in the present work, and
experiment, the levels are $2s^2p^5 \ \ 
(^2P^o_{1/2,3/2})$ and $2s2p^6 \ \ (^2S_{1/2})$. 
The computed target level energies
agree with the observed ones to $<$ 1\%, and the oscillator strengths
agree with those tabulated by NIST (http://physics.nist.gov) to $<$ 5\%. 

We consider photorecombination cross sections for 359 levels of Fe~XVII. 
Total
angular symmetries with J $\leq$ 7 (odd and even), and levels with
$\nu \leq$ 10.0 (($\nu$ is the effective quantum number) 
are included in the photoionization calculations. The photorecombination 
cross sections are obtained via detailed balance using radiatively damped
photoionization cross sections (Pradhan and Zhang 1997). 
Resonances with higher symmetries make negligible contribution. For $10
< \nu \leq \infty$ the calculations are carried out by extending the Bell and
Seaton (1986) theory of DR. As the DR calculations are an
extension of the electron scattering calculations (e.g. NP94,Z99), the
DR and the electron impact cross sections are obtained in a self
consistent manner satisfying the unitarity of the extended electron-photon
S-matrix.

\section{RESULTS}

   Fig. 1a shows the total unified recombination cross section 
$\sigma_{RC}$ for Fe~XVII. It shows photorecombination 
into all 359 levels of Fe~XVII with $\nu \leq 10$ (E $\approx$ 90 eV), 
and DR for $10 < \nu < \infty$ up to E = 131.95 eV, the $^2S_{1/2}$
threshold. 
A rydberg series of resonances converges on to the first excited
level $^2P^o_{1/2}$ at 12.7 eV, followed by the stronger series $n \geq 6$ 
on the $^2S_{1/2}$ level at 131.95 eV. 
It is evident from Fig. 1a that the resonances complexes are
highly resolved, the background contribution in the DR region in
negligible. The resolution in $\sigma_{RC}$ is remarkable given that
it entails extremely detailed photoionization cross sections of hundreds
of bound levels of Fe~XVII. The cross sections have been radiatively
damped; the effect has been found to be much smaller than
for H- and He-like ions (Pradhan and Zhang 1997, Z99). 
The resonance structures are resolved to convergence in the final rates.
This required calculations at extremely fine energy meshes up to $10^{-4}$
eV for individual resonance complexes up to n = 10. 

The nearly fully resolved cross sections exhibit considerably more detail 
than the beam-averaged cross sections in the experiment on the
Heidelberg heavy-ion storage ring (Savin \etal 1999). For the
illustrative comparison in Fig. 1b we 
convolve the theoretical $ v \times \sigma_{RC} $ with a Gaussian
of FWHM 0.020 eV. As
the DR high-n resonances are extremly narrow we use
 an analytic average over the DR cross sections to
obtain the rate coefficients 
in the region $10 < \nu \leq \infty$ below the $2s2p^6 (^2S_{1/2})$ threshold.
At low energies as E $\rightarrow 0$, the $<v * \sigma_{RC}>$ in Fig. 1b
are somewhat lower than the experimental values since the former 
include the non-resonant background up to n $\leq 10$, J $\leq 7$. 
However, the resonance
complexes n = 18 - 20 (and higher) in the near-threshold region have 
been resolved (not shown for
brevity), as in the inset in the lower panel showing experimental results.
In the region E $<$ 12.7 eV, the
non-resonant RR-type contribution dominates the resonant DR-type
contribution (see Fig. 2a and related discussion).
 Although the Gaussian tends to accentuate the peaks rather more
sharply, the agreement in
resonance heights, positions, and shapes of the n = 6 - 10 complexes appears
generally quite good (different
values of FWHM up to 0.050 eV produce little basic change).
The experimental beam shape is simulated by a "flattened" Maxwellian
function with velocity components transverse and parallel to the beam
(e.g. Kilgus \etal 1992), which is then used to fit and extract 
`resonance strengths' (Savin \etal 1999). 
While these may be compared with
theory, a more precise comparison including the non-resonant
background is now possible with the unified cross sections
since the experiment also measures the same. But
because the theoretical results are more detailed, and
owing to extremely narrow widths of resonances, the precise beam
shape function needs to be used for convolution as in experiment.
However, the resonance strengths may be compared independent of the beam
shape function. For example, we find that the integrated cross section
$\sigma_{RC}$ for the n = 7 complex is  350.7 ($10^{-21} cm^2 eV$), compared to
to the MCDF value 335.7 and experimental value of 412.0 $\pm$ 8.1. More
detailed comparisons will be presented in a later report.

 As mentioned earlier, the peak of the DR cross section is theoretically
equal to the threshold electron impact excitation cross section for the
associated dipole core transition.
 The computed DR collision strength at the $^2S_{1/2}$ threshold is 0.27, in
agreement with the electron impact excitation collision strength 
(Berrington and Pelan (2000) obtain 0.2882).
 The DR peak in Fig. 1b is shown with the experimentally determined
field ionization cutoff at $n_{cut} \approx$ 124; the $<v \times
\sigma>$ value is 3.87, compared to 4.18 at the theoretical limit at 
$n = \infty$, a difference of about 8\%, indicating the degree of 
field ionization of the DR peak in the ion storage ring.
 
 Of practical interest in astrophysical models is the total
(e~+~ion) recombination rate coefficient $\alpha_R$(T), including resonant and
non-resonant (RR + DR) contributions, at all temperatures of ionic
abundance. We present the maxwellian averaged $\alpha_R$(T) in Fig.2, using
the computed (unconvolved) $\sigma_{RC}$. Fig. 2a
presents a partial $\alpha_R$(T), 
including the unified $\sigma_{RC}$ (Fig. 1),
compared with the experimental DR results and the multi-configuration 
Dirac-Fock (MCDF) results
(Savin \etal tabulate only the DR rates). 
The unified $\alpha_R$(T) are significantly higher than the experimental 
DR-only rates because the non-resonant (RR-type) contribution
monotically rising towards low-T as $E \rightarrow 0$ 
is included, together with 
the DR bump around 0.4 eV due to the $^2P^o_{1/2} \ n \ \ell$ resonances.
The experimental recombination cross sections also reflect these
two features. The MCDF results are seen to underestimate DR.

Finally, Fig. 2b  presents the total unified
$\alpha_R(T)$ for Fe~XVII (solid curve), using the calculated (unconvolved) cross sections, 
and including a "top-up" non-resonant background contribution for 
recombination into rydberg levels $11 \leq n \leq \infty$
computed in the hydrogenic approximation (e.g. N99a,b). In order to
compare precisely with experimental DR results, we fit and add to it the
non-resonant background contribution from the present total
$\sigma_{RC}$ (Fig. 1a). The difference of $\sim$ 20\% 
between the unified $\alpha_R$  and this sum (`Expt(DR) + non-resonant') 
is then almost entirely due to DR. Also shown is the sum of
experimental DR  and the theoretical RR rate from Arnaud and Raymond (1992).
Although a small contribution may still
be unaccounted for, owing to resolution or high-partial waves, 
the  unified and the experimental results are within 
theoretical or experimental uncertainities.

 We also calculated DR via the forbidden $^2P^o_{3/2} - ^2P^o_{1/2}$
M1 core transition in Fe~XVIII. 
Its contribution is negligible, although
it may be of some consequence in other heavy, highly charged ions
(Pradhan 1983) since A(E2,M1) increases as much higher powers of Z than A(E1).

\section{DISCUSSION AND CONCLUSION}

 Astrophysical X-ray plasmas may be photoionized and/or collisionally
ionized. Modeling of X-ray sources therefore requires total electronic 
recombination cross sections and rates at a wide range of energies and 
temperatures. To that end, we 
calibrate the unified theoretical and experimental data for recombination to 
Fe~XVII against each other; we find good agreement in the low-energy 
region accessible 
experimentally. All other previous unified BPRM calculations have also
shown very good agreement with experiments for few-electron recombined ions 
C~IV, C~V, O~VII (Z99), Ar~XIV (Zhang and Pradhan 1997), and 
Fe~XXIV (Pradhan and Zhang 1997).
Unified recombination cross sections including resonant and
non-resonant processes (RR and DR) may therefore be computed for most
astrophysially abundant elements.

 In future, given the accuracy and resolution of the unified cross sections, 
and possible partial delineation according to n,$\ell$, and J,
an exact comparison with experimental measurements might yield
precise information on: (i) the missing background non-resonant (RR)
contribution near threshold as $E \rightarrow 0$,
into high rydberg levels, (ii) extraneous background
contribution such as due to charge transfer, and (iii) high rydberg resonances
contributing to the DR peak, and a more accurate field ionization cut-off 
than the approximate formulae heretofore employed. 

 Whereas the detailed comparison with experiment (Fig. 1) 
for low-energy $\Delta n = 0$ 
(E $\leq$ 140 eV) recombination is a useful test of accuracy, 
and relevant to low-T photoionized sources,
 collisionally ionized
(coronal) high-T sources require total recombination 
rate coefficients up to T = $10^8$ K (Arnaud and Raymond
1982). That, in turn, requires recombination cross sections up to
E $\approx$ 1 keV or higher. In fact, we may predict that the next DR
peak at about 800 eV due to the n = 3 levels will be much bigger
than at $^2S_{1/2}$ (Fig. 1), since there are several strong dipole 
transitions with
A-values about 1-2 orders of magnitude higher than $A(^2S_{1/2}
\rightarrow ^2P^o_{3/2,1/2})$. The first set of the
these  resonances begins at just above 400 eV, close to the temperature
of maximum abundance of Fe~XVII in coronal equilibrium. 
Correspondingly there will be a higher peak in the recombination 
rate than the one around 100 eV (Fig. 2). The $\Delta n = 1$ DR
therefore is expected to contribute more to total electronic
recombination of (e~+~Fe~XVIII) $\rightarrow$ Fe~XVII 
than the $\Delta n = 0$ core transitions. 
The higher energy unified recombination calculations are in progress.

 One of the main points is that since photoionization and recombination 
are treated as inverse processes with the same eigenfuncation expansion 
for the core ion (Eq. 1), the same set of resonances and non-resonant background
are included in both processes - an essential requirement for 
self-consistency in photoionization equilibrium, as expressed by,
\begin{equation}
  \int_{\nu_0}^{\infty} \frac{4 \pi J_{\nu}}{h\nu} N(X^{z})
\sigma_{PI}(\nu,X^{z}) d\nu = \sum_j N_e N(X^{z+1}) \alpha_R(X_j^{z};T),
\end{equation}
 where $\sigma_{PI}$ is the photoionization cross section and $J_{\nu}$
the radiation flux. In general, the sum on the RHS of Eq. (2) 
extends over the infinite number of recombined bound levels. 
Now if the $\alpha_R$ on the RHS is
subdivided into non-resonant RR and resonant DR, as in existing
photoionization models, then fundamental inconsistencies result. The RR
rate is suppoesedly derived from non-resonant photoionization cross
sections. But as we see from Fig. 1, the near threshold region is
dominated by resonances and a rising non-resonant
background. Therefore, photoionization calculations must also be 
carried out including
the same resonances (c.f. the Opacity Project work; 
Seaton \etal 1994, and references therein). There is however the 
issue of accurate
resonance positions, which may be ameliorated by (a) averaging over 
the radiation field on the LHS of Eq. (2), and (b) pre-averaging over
resonances in photoionization cross sections (Bautista \etal 1998).
 The  further sub-division of DR into $\Delta n = 0$ rates (as, for
example, derived experimentally by Savin \etal 1999)
appropriate for only for low-T plasmas, and $\Delta n > 0$ DR needed in
high-T plasmas, implies that the recombination rates must be obtained
for (RR + DR ($\Delta n = 0$) + DR ($\Delta n > 0$)), generally using different
approximations and possibly valid in different temperature regimes. 
These problems: (i) inconsistent photoionization and recombination,
(ii)unphysical division of RR and DR, and (iii) low and high energy DR 
in different but overlapping energy(temperature) ranges, may be overcome with
the unified method for electronic recombination,
and corresponding photoionization cross sections, to
enable a self-consistent treatment of photoionization
and recombination in X-ray photoionized sources.

\acknowledgments

 We would like to thank Dr. Werner Eissner for invaluable assistance with
the BPRM codes.
 This work was partially supported by the NSF and the NASA
Astrophysical Theory Program.

\newpage

\def\r{\leftskip10pt \parindent-10pt \parskip0pt}
\def\apj{ApJ}
\def\apjs{ApJS}
\def\apjl{ApJL}
\def\aj{Astron. J}
\def\cpc{{Comput. Phys. Commun.}\ }
\def\pasp{Pub. Astron. Soc. Pacific}
\def\mn{MNRAS}
\def\aa{A\&A}
\def\aasup{A\&A Suppl.}
\def\baas{Bull. Amer. Astron. Soc.}
\def\jqsrt{J. Quant. Spectrosc. Radiat. Transfer}
\def\jpb{Journal of Physics B: Atom. Molec. Opt. Phys.}
\def\pra{Physical Review A}
\def\adndt{Atomic Data And Nuclear Data Tables}


\newpage

\begin{figure}
\centering
\psfig{figure=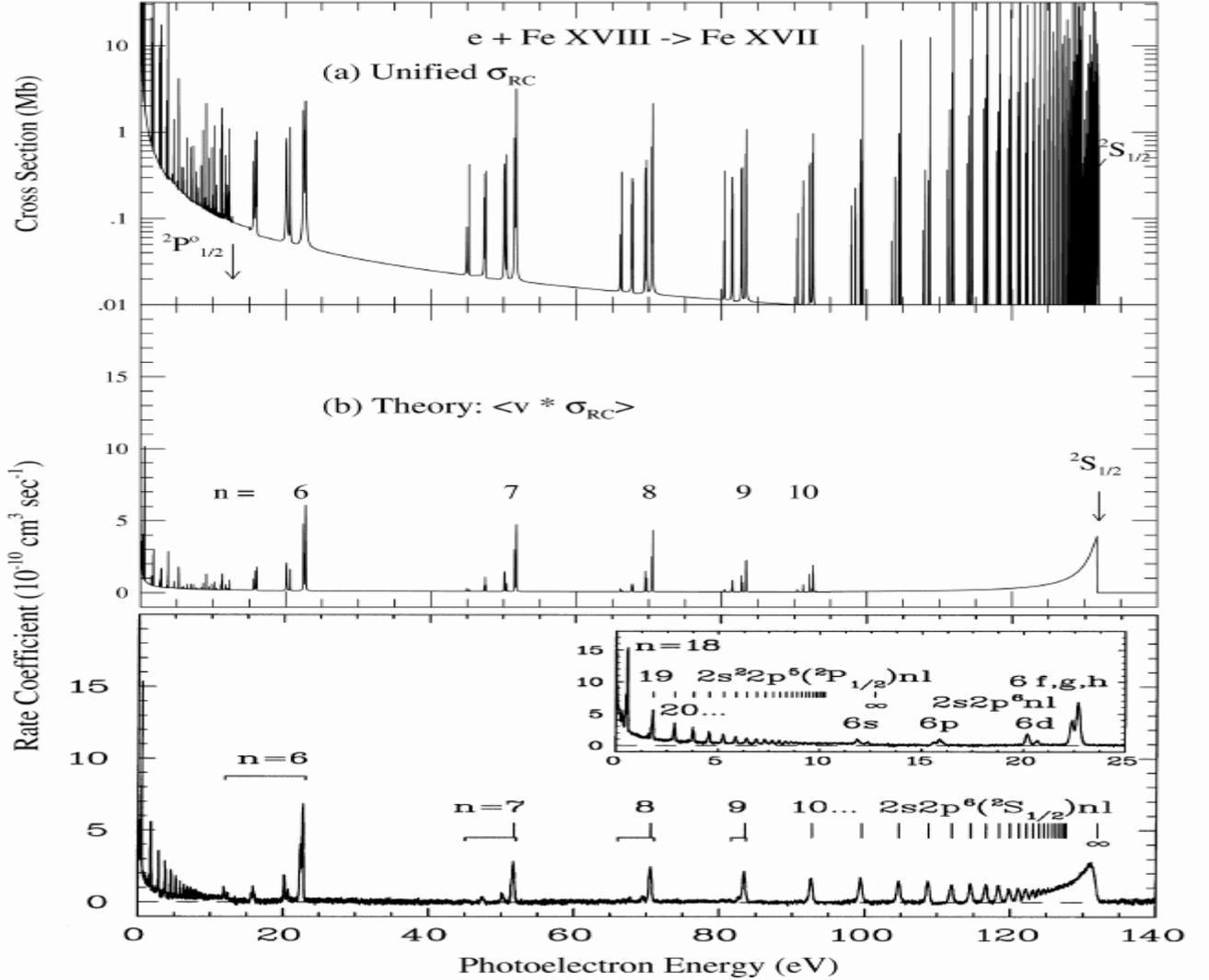,height=15.0cm,width=18.0cm}
\caption[f1.ps]{Unified recombination cross section $\sigma_{RC}$
for Fe~XVII: (a) Narrow rydberg resonances
coverging to the $2s^22p^5 \ (^2P^o_{1/2})$ threshold at 12.7 eV, and
the n-complexes belonging to the $2s2p^6 \ (^2S_{1/2})$ threshold are
delineated. The high-n DR resonances are shown in the region E $>$
90 eV. (b) theoretical rate coefficients $<v \times \sigma_{RC}>$ convolved
over a gaussian beam shape, compared to experimental data in the 
lower panel (Savin \etal 1999). The theoretical DR resonances with 
$10 < \nu \leq
\infty$ below the $^2S_{1/2}$ threshold are analytically averaged.} 
\end{figure}

\begin{figure}
\centering
\psfig{figure=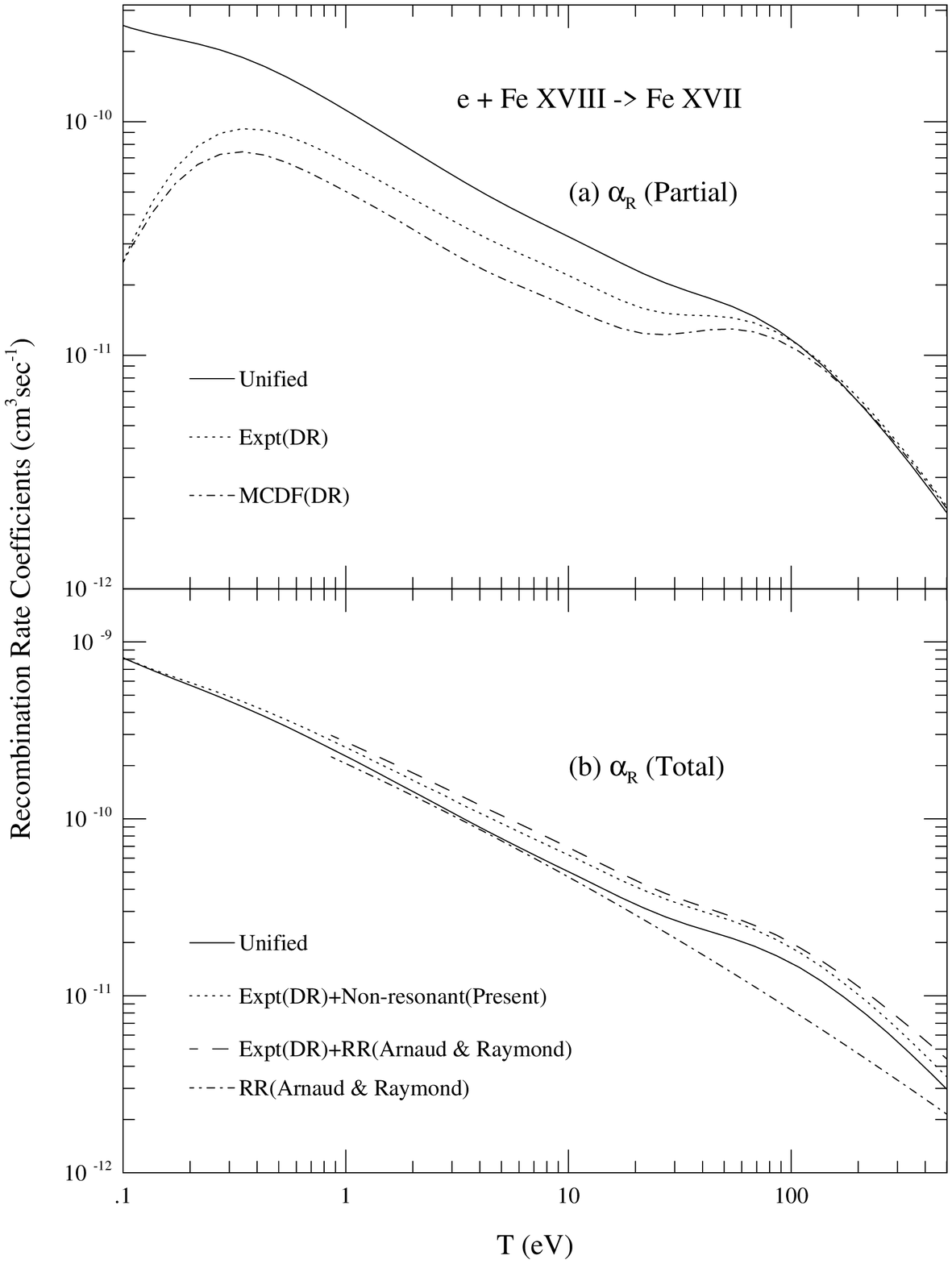,height=15.0cm,width=18.0cm}
\caption[f2.eps]{Unified recombination rate coefficients
$\alpha_R$(T) using computed $\sigma_{RC}$:
(a) partial $\alpha_R$ corresponding to in Fig. 1, compared
with the experimental DR-only rates and  
multi-configuration Dirac-Fock DR rates in Savin \etal (1999).   
The unified $\alpha_R$(T)
are significantly higher since they include the non-resonant
background (including the low-T DR bump at $\sim$ 0.4 eV);
(b) the total $\alpha_R$(T) compared with the sum
of experiment (DR) and the present non-resonant background (RR-type).
The agreement is $\sim$ 20\%. Comparison with the sum of experiment (DR)
and RR (Arnaud and Raymond) is also shown.}
\end{figure}


\begin{thebibliography}{}
\bibitem{} Arnaud, M. \& Rothenflug, D. 1985, A\&AS , 60, 425
\bibitem{} Bautista, M.A., Romano, P., \& Pradhan, A.K. 1998, \apjs, 118, 259
\bibitem{} Bell, R.H. \& Seaton, M.J. 1985, \jpb, 18, 1589
\bibitem{} Berrington, K.A., Eissner, W., Norrington, P.H. 1995,
Comput. Phys. Commun. 92, 290
\bibitem{} Berrington, K.A. \& J. Pelan 2000, \aasup (submitted)
\bibitem{} Brickhouse, N., Raymond, J.C., \& Smith, B.W. 1995, ApJS, 97,
551
\bibitem{} Burke, P.G., Hibbert, A. \& Robb, D. 1971, \jpb, 18, 1589
\bibitem{} Canizares, C.R. \etal in {\it Atomic data needs in X-ray astronomy}
 2000, Eds. M.A. Bautista, T. R. Kallman \& A.K. Pradhan, 
NASA Publications, NASA/CP-2000-209968
(www.heasarc.gsfc.nasa.gov/docs/heasarc/atomic/proceed.html)
\bibitem{} Eissner, W., Jones, M., \& Nussbaumer, H. 1974, \cpc, 8, 270
\bibitem{} Fabian, A.C., Mushotzky, R.F., Nulsen, P.E.J. \& Peterson,
J.R. \mn (in press, astro-ph/0010509)
\bibitem{} Hummer D.G., Berrington K.A., 
Eissner W., Pradhan A.K, Saraph H.E., \& Tully J.A., 1993, \aa, 279, 298
\bibitem{} Kallman, T. 1995, in {\it Atomic Processes in Plasmas}, 
AIP Press, New York
(http://heasarc.gsfc.nasa.gov/docs/heasarc/atomic/proceed.html)
\bibitem{} Kilgus, G., Habs, D., Schwalm, D., Wolf, A., Badnell, N.R. \&
M\"{u}ller, A. 1992, \pra, 46, 5730
\bibitem{} Mannervik, S., Asp, S, Brostr\"{o}m, L, DeWitt, D.R.,
Lidberg, L., Schuch, R. \& Chung, K.T. 1996, \pra, 55, 1810
\bibitem{} Nahar, S.N. \& Pradhan, A.K. 1994, \pra, 49, 1816 (NP94)
\bibitem{} Nahar, S.N., Pradhan, A.K. \& Zhang, H.L. 2000a, \apjs (in
press astro-ph/0003411); 2000b, \apjs (in press, astro-ph/0008023)
(N99a,b)
\bibitem{} Ogle, P.M., Marshall, H.L., Lee, J.C. \& Canizares, C.R. 2000
(in press, astro-ph/0010314)
\bibitem{} Predehl, P. \etal in {\it Atomic data needs in X-ray astronomy}
 2000, Eds. M.A. Bautista, T. R. Kallman \& A.K. Pradhan, 
NASA Publications, NASA CP-2000-209968
\bibitem{} Pradhan, A.K. \& Zhang, H. L. 1997, J. Phys. B 30, L571
\bibitem{} Pradhan, A.K. 1983, \pra, 28, 2128
\bibitem{} Savin, D.W, Bartsch, T., Chen, M.H., Kahn, S.M., Liedahl,
D.A., Linkemann, J., M\"{u}ller, A., Schippers, S., Schmitt, M.,
Schwalm, D. \& Wolf, A. 1997, \apjl, 489, L115
\bibitem{} Savin, D.W., Kahn, S.M., 
Linkemann, J., Saghiri, A.A., Schmitt, M., Grieser, M., Repnow, R.,
Schwalm, D., Wolf, A., Bartsch, T., Brandau, C., Hoffknecht, A.,  
M\"{u}ller, A., Schippers, S., Chen, M. \& Badnell, N.R. 1999, \apjs,
123, 687
\bibitem{} Schulz, N.S., Canizares, C.R., Husenmoerder, D, \& Lee, J.C.
\apjl (in press, astro-ph/0010310)
\bibitem{} Scott, N.S. \& Taylor, K.T. 1982 \cpc, 25, 347
\bibitem{} Seaton, M.J., Yu Yan, Mihalas, D., \& Pradhan, A.K. 1994,
\mn, 266, 805
\bibitem{} Zhang, H.L. \& Pradhan, A.K. 1997, \prl, 78, 195
\bibitem{} Zhang, H.L., Nahar, S.N. \& Pradhan, A.K. 1999, J. Phys. B,
32, 1459 (Z99)

\end{thebibliography}
\end{document}